# High Frequency Radio Network Simulation Using OMNeT++

Jeffery Weston

Eric Koski

RF Communications Division
Harris Corporation
Rochester, NY, USA

*Abstract*— **Harris Corporation has an interest in making HF radios a suitable medium for wireless information networks using standard Internet protocols. Although HF radio links have many unique characteristics, HF wireless subnets can be subject to many of the same traffic flow characteristics and topologies as existing line-of-sight (LOS) radio networks, giving rise to similar issues (media access, connectivity, routing) which lend themselves to investigation through simulation. Accordingly, we have undertaken to develop efficient, high-fidelity simulations of various aspects of HF radio communications and networking using the OMNeT++ framework. Essential aspects of these simulations include HF channel models simulating relevant channel attributes such as Signal to Noise Ratio, multipath, and Doppler spread; a calibrated physical layer model reproducing the error statistics (including burst error distributions) of the MIL-STD-188-110B/C HF modem waveforms, both narrowband (3 kHz) and wideband (up to 24 kHz) on the simulated HF channels; a model of the NATO STANAG 5066 data link protocol; and integration of these models with the OMNeT++ network simulation framework and its INET library of Internet protocol models. This simulation is used to evaluate the impacts of different STANAG 5066 configuration settings on TCP network performance, and to evaluate strategies for optimizing throughput over HF links using TCP Performance Enhancing Proxy (PEP) techniques.**

## I. INTRODUCTION

High Frequency (HF) radio has long been considered to have limited potential for data communications [1], due to highly variable propagation conditions, high error rates, long and variable latency, and data rates that (in narrowband channels) do not exceed and frequently cannot reach 9600 bits per second (BPS).

However, HF has several advantages that can offset the disadvantages. Foremost is the fact that HF radio signals can reliably propagate beyond line-of-sight (BLOS, also referred to as skywave), often around the world, with no dependence on a fixed network infrastructure.

BLOS communications over long distances can typically be achieved only within a 3 to 6 MHz wide band of the overall HF spectrum for any given time and station locations. Which band can be used depends on the time of day, season, and solar radiation characteristics that can be predicted *a priori* with only partial success. For this reason, most critical HF communications systems rely on near-real-time propagation sensing and frequency adaptation through the use of Automatic Link Establishment (ALE) techniques.

HF also exhibits interesting groundwave propagation mode which can result in successful propagation over distances somewhat longer than line-of-sight. HF groundwave propagation requires an entirely separate propagation model, and is the subject of distinct research and study.

## II. NETWORK MODEL OVERVIEW

It is useful to view a network model as comprising three parts [5]:

- Traffic model – the nature and behavior of the traffic source and sink, including arrival rate, "burstiness", Quality of Service (QoS) attributes, etc.

- Bearer model – the parts of the network that are actually used to transport the required traffic.

- Channel model – the medium across which the traffic is delivered.

We will discuss each of these sub-models in reverse order, since the unique needs of the channel model drive the other sub-model requirements. Refer to Figure 1 for a graphical view of the module layout and interconnections.

### A. Channel Model

The physical factors affecting HF propagation are materially different than those affecting LOS radio propagation. The primary considerations in LOS radio propagation are free space loss (inverse-square distance loss), ground reflection (as represented by two-ray propagation models), shadowing (by terrain or artificial features) and diffraction. Most of these are of secondary importance to HF skywave propagation, where the primary factors are ionospheric refraction and absorption spread due to simultaneous propagation paths with differential delay, multipath, and comparatively large Doppler spreads. HF groundwave propagation has its own set of unique considerations, including ground conductivity and permittivity, as well as shadowing and diffraction by terrain features.

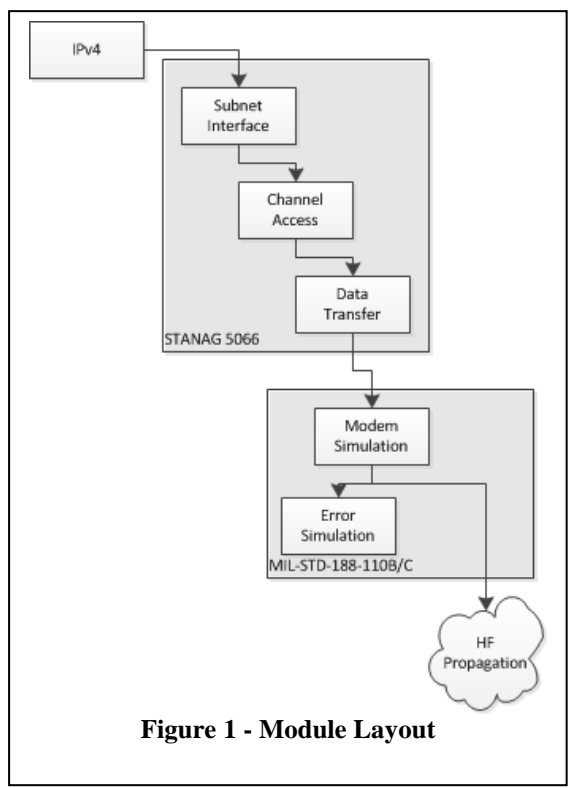

**Figure 1 - Module Layout**

The standard model for the HF channel has been the Watterson model with standard channel conditions defined by ITU-R Recommendation Rec. F.1487 [6]. DSP-based simulators that implement this model at the baseband level have existed for many years, and have been used to test HF modem designs. In general, these simulators use fixed values for Signal to Noise Radio (SNR), Doppler spread, and multipath delay, and simulate the signal perturbations of the HF channel quite effectively over periods of up to a few seconds [7]. However, these simulators do not model variations in signal quality that have been observed with periods of more than a few seconds to minutes.

While it would not be impossible to integrate a software implementation of this model into our simulation, the extensive signal processing computations required make them too slow for network simulations.

It is also important to note that the data rates and statistical nature of HF channel variations require much longer run times to yield valid results. A useful simulation thus must not only generate valid results, but must generate those results in much less than real time.

We have developed a software-based HF channel simulator that benefits from state-of-practice HF channel models and provides narrowband and wideband simulation capability. Modem performance statistics obtained in conjunction with this channel simulator have been used to define and calibrate the physical layer of our bearer model, as is further explained in the next section.

### B. Bearer Model

The bearer part of the model consists of a modem simulation component, a data link protocol simulation component, and the internet protocol stack.

#### 1) Modem Simulation Component

The MIL-STD-188-110B/C [3] modem simulation consists of a single simulation module which implements the functional module simulation, plus a separate C++ module which provides the propagation error simulation code (this division exists both for division of labor reasons and to allow the error simulation code to also operate as a standalone application for other uses). While it may seem more correct to place the error simulation in the HF propagation module, in fact it is more correct to say that the errors are made in the modem as it receives and decodes the incoming signal, and the pattern of errors is a function of the modem settings and the received signal plus interference.

A fundamental challenge to implementing a modem simulation is the signal processing computations required, which would adversely affect the ability to run simulations in an acceptable period of time.

The solution implemented for this simulation is to run the software-based channel and modem simulations offline to generate lengthy error traces (binary sequences containing a 1 for each simulated modem bit-error). These traces were then analyzed to determine cumulative distributions for the lengths of error-free runs and mixed runs containing errors, for each combination of channel conditions (SNR, Doppler shift, and multipath spread) and modem parameters (data rate and interleaver depth). These cumulative distributions, in table form, are then used (in the simulation framework) by a software component referred to as the 'Erracle' (Error + Oracle) to generate simulated errors having equivalent run-length distributions, so as to accurately model the 'burstiness' of the error sequences. In this manner the Erracle is capable of generating bit errors with the appropriate distribution excess of $10^7$ bits per second.

The modem simulation is also responsible for determining the time required to transmit a given block of data, based on the data rate and other modem settings. Accurate transmit time information is naturally critical to ensuring that the simulation results are accurate; our modem model provides this timing information based on detailed knowledge of the modem transmission format as defined by MIL-STD-188-110B/C.

### 2) Data Link Protocol (STANAG 5066) Component

The STANAG 5066 [4] Simulation is broken into three submodules, corresponding to the three required functional capabilities of the protocol, SubnetInterface, ChannelAccess, and DataTransfer.

SubnetInterface is responsible for accepting IP packets from the IPv4 module at the source side of the transfer and delivering those packets at the destination side of the transfer. To simplify the lower levels of the simulation, the actual packet received is transferred via a "virtual" interface (that is, an interface that has no analog in STANAG 5066) to the destination SubnetInterface. The packet is held there until the data frame(s) that comprise the packet have been declared delivered by the lower layers of the simulation. Since the entire packet has been cached in the destination SubnetInterface, it is not necessary to actually fragment and re-assemble the packet in the lower levels, dramatically simplifying the simulation without sacrificing any simulation fidelity.

The SubnetInterface implementation is also derived into two C++ subclasses, designed to allow the exploration of different capabilities. A "Saturation" subclass ignores the network interface entirely and simply generates enough test packets to keep the interface operating at peak efficiency, to allow us to measure optimal bulk traffic throughput. This removes TCP and control messages, as well as making certain that sufficient traffic is always available. Theoretically, throughput measured in this mode should be a benchmark against which network throughput values can be compared, and can be used to calibrate and validate the model.

A second subclass of SubnetInterface is Acceleration, which implements the capabilities described in section III. Many parameters are exposed at this level which allows us to run simulations designed to optimize network traffic flow.

ChannelAccess is responsible for creating the HF link over which the transfer will take place. Currently, we make the assumption that the radios are already linked on a common frequency, so this module is a stub.

DataTransfer is responsible for simulating the fragmentation of packets into STANAG 5066 data frames, scheduling the transmission of those frames, verifying the proper reception of the frames, and re-assembling them into the original packets. For purposes of simulation, frames are simulated as structures that define the size and location of the data the frame represents. As noted above, the packet data is not actually fragmented or reassembled in the simulation, a substantial simplification that does not reduce the fidelity of the simulation.

### 3) Internet Protocol Stack

To properly simulate the network, it is necessary to simulate the internet protocol stack. In the case of the HF network, it is desirable to intercept packets at the IP level, but the behavior of the higher levels must also be taken into account.

The INET framework of OMNeT++ is ideal for our purposes, in that it includes detailed and tested simulations of TCP (in multiple variations), UDP, IP, HTTP, VOIP, and a variety of standard wireless protocols. We are indebted to the OMNeT++ community for the large and thorough base of existing work, in that it allows us to concentrate our efforts on the aspects of the simulation that are unique to HF radio.

### C. Traffic Model

For our initial development, the traffic model is essentially stubbed out. We are using the TCPApp simulation component of the INET framework to generate TCP traffic of varying sizes in order to drive the simulation to produce results.

## III. PERFORMANCE ENHANCING PROXY OVERVIEW

As expected, TCP over HF tends to work successfully only under a limited set of channel conditions. Furthermore, where it does work, it often provides limited throughput. To provide acceptable throughput, it is desirable to manage the interface a more active manner.

A TCP Performance Enhancing Proxy (PEP) [2] is used to isolate TCP from links over which it cannot be supported directly (due to error rate, latency, or other issues). The PEP intercepts IP packets as they are being routed, and masquerades as the TCP final destination to carry on the transfer with the origin TCP node. The packets are then transported through the problematic link (typically using a different protocol specifically adapted to the physical channel), where the receiving end of the PEP establishes a separate TCP connection with the real destination node, completing the transfer. TCP at both the source and destination nodes are thus isolated from the difficulties of the intermediate channel, and are provided with the illusion of a "normal" direct connection.

In the case of narrowband HF links, the performance issues to be overcome are:

1. High error rates – Under simulation, packet error rates as high as 4% have been measured on links that were able to carry TCP traffic.

2. High latency – The STANAG 5066 standard recommends transmitting for up to 120 seconds, then waiting for the acknowledgment, before transmitting again. This implies normal packet Round-Trip Times (RTT) of over 2 minutes.

3. High latency jitter – When a packet is received in error, it must be retransmitted with the next transmission. Therefore, the latency of that particular packet would typically exceed 4 minutes. In addition, as error rates rise, HF data link protocols often respond by lowering the data rate for increased robustness, further increasing the RTT.

To combat issues with the link, the PEP may locally generate TCP acknowledgments (ACKs) prior to actually delivering the data, aggregate ACKs to limit congestion, or take other actions that are required to present an interface to the source or (less often) destination TCP implementation that conforms to expectations and allows the transfer to continue.

## IV. PRELIMINARY RESULTS

The first results of this simulation were used to calibrate the model versus existing STANAG-5066 test data, to ensure that the core of the model was accurately simulating the expected behavior of the data link protocol and HF channel. Table 1 shows ideal (no-error) throughput for both the actual system and the a calculated idea throughput for various data rates, and demonstrates that the simulation does in fact accurately model STANAG-5066 at this level.

| Table 1 - Model Throughput Calibration | | | |
|---|---|---|---|
| Data Rate (bps) | Model Throughput | Calculated Throughput | Percent Difference |
| 75 | 52 | 57 | -8.8% |
| 150 | 115 | 113 | 1.8% |
| 300 | 228 | 229 | -0.4% |
| 600 | 465 | 456 | 2.0% |
| 1200 | 890 | 912 | -2.4% |
| 2400 | 1760 | 1803 | -2.4% |
| 3200 | 2597 | 2543 | 2.1% |
| 4800 | 3812 | 3717 | 2.6% |
| 6400 | 4711 | 4536 | 3.9% |
| 8000 | 5641 | 5476 | 3.0% |
| 9600 | 6380 | 6202 | 2.9% |

The next simulation series was intended to demonstrate whether and how well TCP data could be carried over HF. As previously noted, reliable and usable TCP over HF has historically been very hard to achieve with reasonable throughput. We implemented a PEP as described in section III. In addition, we have also been experimenting with various proprietary enhancements to the standard PEP strategy, which we have termed Enhanced TCP Acceleration. Figure 2 shows the probability of successful TCP transfer versus SNR for TCP, TCP with Acceleration, and TCP with Enhanced Acceleration.

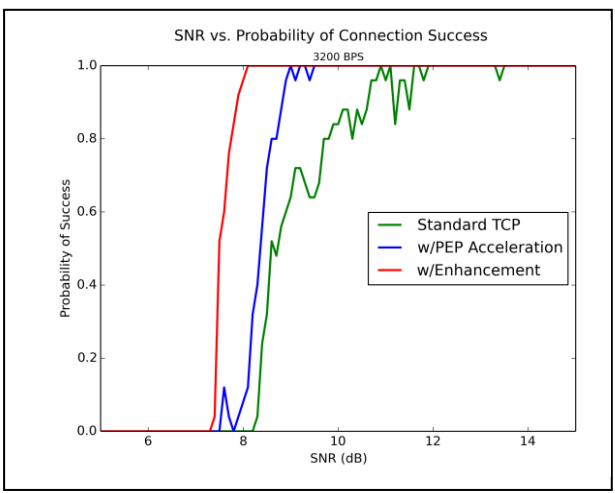

The abruptness of the transitions is to be expected, both due to the fact that we are only measuring whether the transfer completed, and also due to the nature of HF modems in a static average SNR environment. More realistic long-term and medium-term variations in the Modem Model are planned. In addition, it is common to adapt the data rate of the modem to current conditions, which will likely smooth out the probability curve. This strategy will also, however, confer a throughput advantage to these enhancements.

## V. FUTURE WORK

With the first version of the HF Data Network simulation working, a number of opportunities for enhancement and expansion suggest themselves;

Channel Model Improvements: It may be useful to integrate propagation predictions from a propagation model such as VOACAP directly into the simulation, simplifying the scenario configuration process.

Modem Model Improvements: It is also very likely that we will modify the channel model to take advantage of improvements in modeling such as the intermediate and long term variations in SNR described above.

Wideband STANAG 5066: We intend to modify the channel error and modem models to simulate wideband signaling. This will also require modifications to the STANAG 5066 simulation, as some of the protocol settings will not efficiently support the higher data rate transfers supported by wideband. Both the wideband signaling capability and the subsequent required modifications to STANAG 5066 are currently being considered by the NATO standardization committee, and it is anticipated that the ability to simulate these changes will provide substantial input into those discussions.

Channel Access: To realistically simulate actual HF networks, it will be necessary to simulate the process of locating an acceptable channel and creating a link with one or more other nodes on that channel. Initial efforts will simulate existing narrowband ALE protocols; later efforts will include simulation and evaluation of proposed wideband ALE technology.

Advanced Protocols: With the introduction of wideband HF, new avenues of inquiry have opened up regarding data protocols. We have also been studying so-called "Hybrid-ARQ" protocols, in which the modem and data link protocol are more closely intertwined. This allows data transfers to take advantage of diversity combining and code combining techniques to increase reliability and decrease latency. Simulation will provide us with a unique opportunity to study the performance of these techniques prior to their implementation in the target system.

Traffic Model Improvements: We anticipate that improvements to the traffic model will continue into the long term, since it is expected that each potential user environment may have its own unique traffic profiles. In the immediate future we anticipate the development of a number of representative traffic profiles to exercise the simulation in a more realistic manner.

## VI. CONCLUSIONS

This simulation has been implemented in a relatively short period of time, and has already begun to produce useful and interesting results. We anticipate that the usefulness will only increase as we expand the simulation into new areas of inquiry, and presume that the output of this simulation will drive not only the development of future products, but also the development of interoperability standards worldwide.

## VII. ACKNOWLEDGEMENTS

The authors gratefully acknowledge the assistance of John Nieto in the development of the HF channel model. The authors would also like to thank the OMNeT++ community for the development of this toolset and for the documentation and on-line assistance that is available, all of which were of great assistance in the development of this model.